\documentclass[aps,prl,showpacs,twocolumn]{revtex4}

\usepackage{graphicx}
\usepackage{dcolumn}
\usepackage{bm}
\usepackage{color}
\usepackage{amsmath}
\usepackage{amssymb}

\begin{document}
\title{Detection-dependent six-photon NOON state interference}
\author{Rui-Bo Jin$^{1,2}$}
\email{jin@wit.edu.cn}
\author{Mikio Fujiwara$^{1}$,  Ryosuke Shimizu$^{3}$,  Robert J. Collins$^{4}$, Gerald S. Buller$^{4}$,   Taro Yamashita$^{5}$,    Shigehito Miki$^{5}$,   Hirotaka Terai$^{5}$,  Masahiro Takeoka$^{1}$,   and Masahide Sasaki$^{1}$}

\affiliation{$^{1}$National Institute of Information and Communications Technology (NICT), 4-2-1 Nukui-Kitamachi, Koganei, Tokyo 184-8795, Japan}

\affiliation{$^{2}$School of Science and Laboratory of Optical Information Technology, Wuhan Institute of Technology, Wuhan
430205, China}

\affiliation{$^{3}$University of Electro-Communications (UEC), 1-5-1 Chofugaoka, Chofu, Tokyo 182-8585, Japan}

\affiliation{$^{4}$SUPA, Institute of Photonics and Quantum Sciences, School of Engineering and Physical Sciences,  \\ Heriot-Watt University, David Brewster Building, Edinburgh EH14 4AS, United Kingdom}

\affiliation{$^{5}$National Institute of Information and Communications Technology (NICT), 588-2 Iwaoka, Kobe 651-2492, Japan}

\date{\today }

\begin{abstract}
%
NOON state interference (NOON-SI) is a powerful tool to improve the phase sensing precision, and can play an important role in quantum sensing and quantum imaging.
However, most of the previous NOON-SI experiments only investigated the center part of the interference pattern, while the full range of the NOON-SI pattern has not yet been well explored.
In this Letter, we experimentally and theoretically demonstrate up to six-photon NOON-SI and study the properties of the interference patterns over the full range.
The multi-photons were generated at a wavelength of 1584 nm from a PPKTP crystal in a parametric down conversion process.
It was found that the shape, the coherence time and the visibility of the interference patterns were strongly dependent on the detection schemes.
%
%
This experiment can be used for applications which are based on the envelope of the NOON-SI pattern, such as quantum spectroscopy and quantum metrology.
\end{abstract}



\maketitle

\textbf{\emph{Introduction}}
\noindent
Multi-photon entanglement and multi-photon interference are useful nonclassical phenomenon in quantum information applications.
In particular, the so-called NOON state interference (NOON-SI) is a powerful tool to improve the phase sensing precision.
The NOON state is a path-entangled state with  $N$ photons occupying either one of two optical paths: $\frac{1}{{\sqrt 2 }}(\left| {N0} \right\rangle  + \left| {0N} \right\rangle )$  \cite{Boto2000}.
NOON states can perform super-sensitive measurements, therefore, have been widely used in quantum lithography \cite{Boto2000},  quantum metrology \cite{Giovannetti2011}, quantum microscopy \cite{Ono2013, Israel2014},   and  error correction \cite{Bergmann2015}.
Many NOON-SI experiments have been carried out with photon numbers (values of $N$) from two \cite{Edamatsu2002} to three \cite{Mitchell2004}, four \cite{Walther2004, Nagata2007}, five \cite{Afek2010} and six \cite{Xiang2013} at visible wavelengths,  and also at telecom wavelengths \cite{Yabuno2012, Bisht2015}.

All these previous NOON-SI schemes mainly considered the period-based applications and only measured the single-mode interference pattern, which is only the middle portion of the full-range interference patterns. However, the full-range properties of the NOON-SI, e.g., the envelope shape and the coherence time,  are omitted.
Therefore the previous experiments are not enough to fully characterize the NOON-SI, especially for higher photon numbers ($N$$>$2).
In this Letter, we focus on the full range properties of the NOON-SI pattern and experimentally measured an up to six-photon NOON-SI pattern at telecom wavelength, generated by a spontaneous parametric down conversion (SPDC) source.
It was found that the the shape, the coherence time and the visibility of the interference patterns were strongly dependent on the detection schemes.
We also developed a multi-mode theory for NOON-SI, which corresponds well with  the experimental results.
Further, the effect of multi-photon emission in NOON-SI are analyzed in detail.
The experiment and theory in this work are useful for the future envelope-based applications, such as quantum spectroscopy \cite{Dinani2016, Whittaker2015} and quantum metrology \cite{Dowling2008, Taylor2016}

\begin{figure}[tbp]
\centering
\includegraphics[width= 0.48\textwidth]{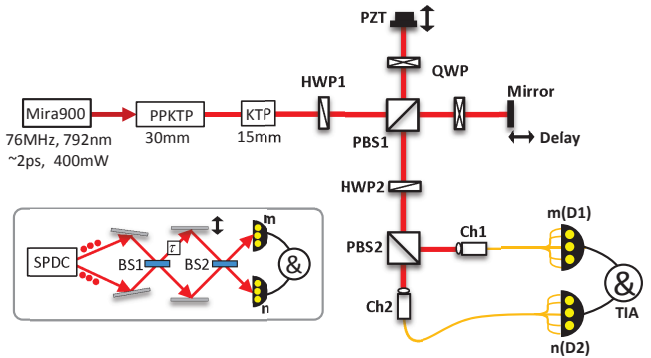}
\caption{(color online)  The experimental setup.  HWP= half wave plate, QWP=quarter wave plate, PZT=piezo-electric linear actuator, PBS=polarization beam splitter, TIA=time interval analyzer.
The inset depicts a standard configuration of the NOON type interference using path-mode.
}
\label{setup}
\end{figure}

\textbf{\emph{Experiment and results}}
\noindent
Our experimental setup for multi-photon NOON-SI is shown in Fig.\,\ref{setup}.
%
%
%
Picosecond laser pulses (pulse repetition frequency = 76 MHz, wavelength = 792 nm, temporal duration $\sim$ 2 ps) from a mode-locked Titanium sapphire laser (Mira900, Coherent Co.) pumped a  30-mm-long PPKTP crystal with a poling period of 46.1 $\mu$m for type-II group-velocity-matched SPDC \cite{Konig2004,Jin2013OE}.
The down converted photons, i.e., the signal and idler photons, had orthogonal polarizations and degenerate wavelengths at 1584 nm (temperature was 32.5\,$^{\circ}$C for the PPKTP crystal).
To compensate for their different group velocities during propagation through the nonlinear crystal, the downconverted biphotons passed through a KTP crystal of half the length of the previous crystal (15 mm).
After the biphotons were mixed on a half wave plate (HWP1, at 22.5\,$^{\circ}$), they were sent to a time scanning system, which was composed of a polarization beam splitter (PBS1), two quarter wave plates (QWP, at 45\,$^{\circ}$) and two mirrors.
One of the mirrors was set on a PZT linear actuator to achieve a scanning step  in the order  of nm, while the other one was on a stepping motor linear actuator to realize a scanning step of $\mu$m.
Then, the biphotons were mixed again on the second half wave plate (HWP2, at 22.5\,$^{\circ}$) and separated by PBS2, before they were collected into two channels of single-mode fibers (Ch1 and Ch2).
Ch1 was connected to a 1$\times$$m$ fiber coupler and each of the $m$ output ports is coupled to a superconducting nanowire single-photon detector (SNSPD) \cite{Miki2007, Miki2013}.
The same conditions were set for Ch2 with a 1$\times$$n$ fiber coupler  and $n$ SNSPDs.
Finally, all the detected events were sent to a time interval analyzer (TIA) for coincidence counting.

\begin{figure*}[tbp]
\centering
\includegraphics[width= 0.95\textwidth]{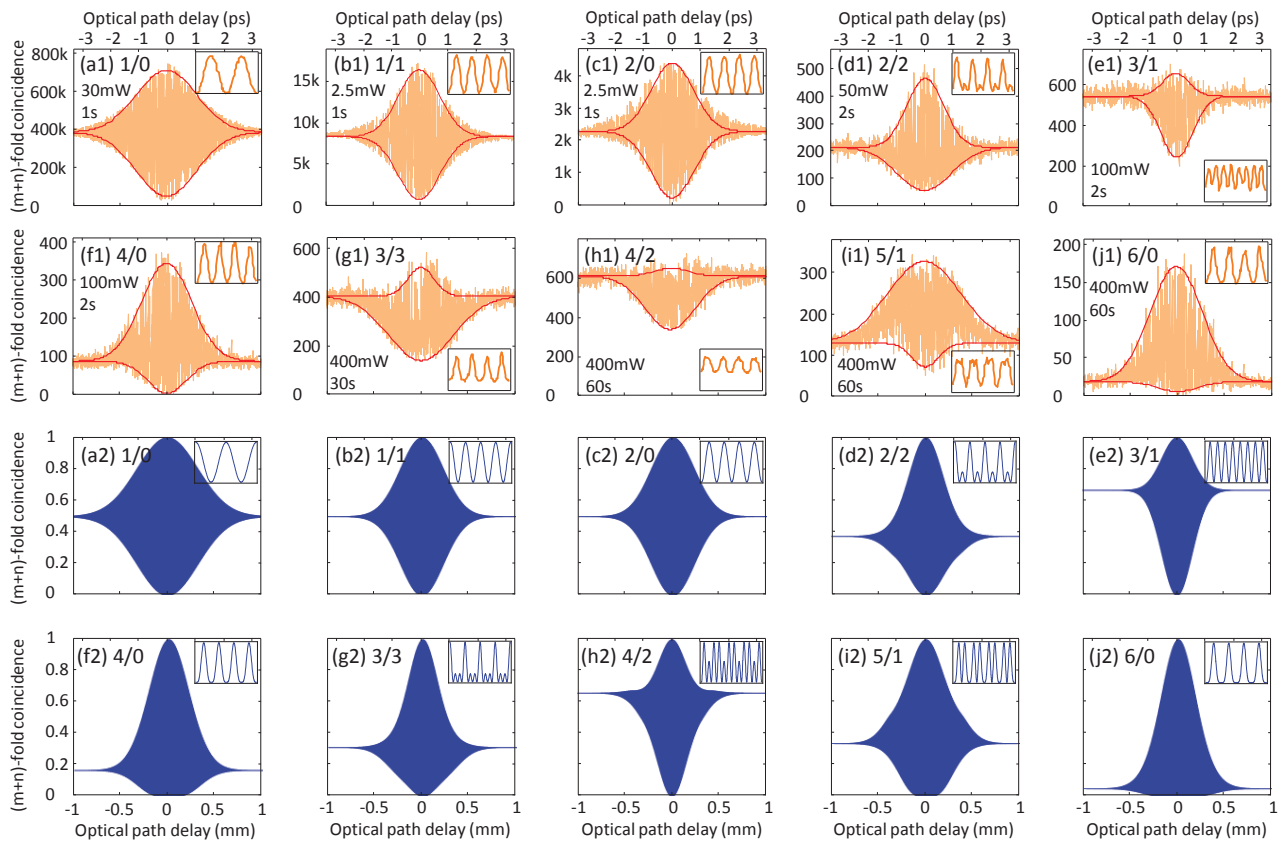}
\caption{ (color online)   Experimental results and numerical simulations of multi-photon NOON-SI under the $m$/$n$ detection schemes.
The first two rows (a1-j1) are the experimental results obtained by scanning a stepping motor for  1000 steps with a step length of 2 $\mu$m.  The parameters on the left side are the pump power (in mW) and accumulation time (in seconds, for each point). The inset in each figure is the interference pattern obtained by scanning a Piezo (PZT) near the zero delay position. The horizontal axis for each inset is phase delay from 0 to 4$\pi$, while the vertical axis is the same as each main figure. The third and fourth rows (a2-j2) are the  corresponding numerical simulations using the theoretical model. Insets in  (a2-j2) are the fine interference patterns near the zero delay position.
}
\label{results}
\end{figure*}

In this experiment, we used six SNSPDs with dark counts less than 1 kHz and detection efficiencies around 70\%  \cite{Jin2015OC}.
$m$ SNSPDs were used for Ch1 and $n$ SNSPDs were for Ch2.
By changing the different fiber couplers, we could carry out different $m/n$ detection schemes.
The 1/1, 2/0, 2/2, 3/1, 4/0, 3/3, 4/2, 5/1 and 6/0 detection results are shown in Fig.\,\ref{results}(a1-j1).
%
%
%
%
%

In Fig.\,\ref{results}(a1-j1), all the data in the figure are presented as raw data without subtracting any background counts.
It is clearly shown  in Fig.\,\ref{results}  that the profile, the coherence time and the visibility of the interference patterns  are dependent on the detection schemes.
Figure\,\ref{results}(a1) shows the one-photon NOON-SI, with only the signal photons input and only one SNSPD for detection. The idler photons are blocked by inserting a PBS after the PPKTP crystal.
To overcome the problem of dark counts in the SNSPD, we utilized a pump power of 30 mW for the one-photon interference.
Figure\,\ref{results}(b1-c1)  shows the two-photon NOON-SI with the detection schemes of $1/1$ and $2/0$.
Figure\,\ref{results}(d1-f1)  are the four-photon NOON-SI patterns with the detection schemes of $2/2$, $3/1$  and $4/0$.
Figure\,\ref{results}(g1-j1)  are the six-photon NOON-SI patterns with the detection schemes of $3/3$, $4/2$, $5/1$ and $6/0$.
To obtain the experimental data in a reasonable time, we adopted a high pump power of 400 mW for the six-photon interference.

We can quantitatively evaluate the interference patterns in Fig.\,\ref{results}(a1-j1)  by using three  different  parameters: the shape, the coherence length and the visibility,  as shown in Table \ref{table1}.
\begin{table*}[tbp]
\centering\begin{tabular}{c|cccccccccc}
\hline \hline
 Detection scheme $ m/n $        &1/0  &1/1    &2/0     &2/2     &3/1   &4/0    &3/3   &4/2  &5/1    &6/0\\
 \hline 
  Profile shape                 & sym.  &sym.   &sym.      &bump  &dip  &bump  &dip  &dip  &bump  &bump \\
  Coherence length (mm)         & 0.75  &  0.53  &  0.53   & 0.46  & 0.40  & 0.63  &  0.81   & 0.62  & 0.92  & 0.65\\
  Coherence time (ps)           & 2.5   &  1.77  &  1.77   & 1.53  & 1.33  & 2.10  &  2.70   & 2.07  & 3.07  & 2.17\\
  Visibility                    & 0.99  & 0.92   & 0.98    & 0.85  & 0.53  &  0.98  & 0.63   & 0.35  & 0.73  & 0.98 \\
  \hline \hline
\end{tabular}
\caption{\label{table1}
Parameters of the experimental patterns in Fig.\,\ref{results}(a1-j1). sym.=symmetric shape.}
\end{table*}
We classify the shape of the interference patterns into three categories: symmetric profile, dip or bump.
The $1/0$, $1/1$ and $2/0$ detection schemes have a symmetric profiles, since the upper envelope and the lower envelope have the same bandwidth.
The $2/2 $, $4/0$, $5/1$   and $6/0$ detection schemes mainly show  bump profiles.
The  $3/1 $, $3/3$ and $4/2$  detection schemes  are dominantly dips.
The coherence length (bandwidth) of bumps are defined as the full-width at half maximum (FWHM) of the upper envelope, while the bandwidth of the dips are defined as the FWHM of the lower envelopes.
It can be noticed that the $3/1$ detection scheme has the smallest bandwidth, while $5/1$ has the largest bandwith.
The coherence time is directly calculated from coherence length by dividing the speed of light.
The visibilities are calculated from the nm scale scanning step data in the inset in Fig.\,\ref{results}(a1-j1).
The visibility of $4/2$ schemes has lowest visibility, however, the $1/0$, $2/0$, $4/0$, $6/0$ always keep high visibilities, even at high pump powers.

\textbf{\emph{Theoretical analysis}}
\noindent
To fully explain the experimental results, we developed multi-mode theory using Schmidt decomposition on the temporal modes of the NOON state \cite{Tichy2011, Ra2013NC, Ra2013PNAS}.
The calculated detection probability $P_{mn}$ for each $m/n$ detection schemes are:
\begin{eqnarray}
  P_{10}  &=&  \frac{1}{2}[1 + \sqrt {I(\tau )} \rm{cos}(\omega \tau )] \\
  P_{11}  &=& \frac{1}{2}[1 + I(\tau )\rm{cos}(2\omega \tau )] \\
  P_{20}  &=& \frac{1}{4}[1 - I(\tau )\rm{cos}(2\omega \tau )] \\
     \nonumber
  P_{22}&=& \frac{1}{{32}}[12 - 4I(\tau ) + 3I(\tau )^2  + 12I(\tau )\rm{cos}(2\omega \tau ) \\
         &&+ 9I(\tau )^2 \rm{cos}(4\omega \tau )] \\
  P_{31} &=& \frac{1}{{16}}[4 - I(\tau )^2  - 3I(\tau )^2 \rm{cos}(4\omega \tau )] \\
     \nonumber
   P_{40} &=& \frac{1}{{64}}[4 + 4I(\tau ) + I(\tau )^2  - 12I(\tau )\rm{cos}(2\omega \tau ) \\
      &&  + 3I(\tau )^2 \rm{cos}(4\omega \tau )]
\end{eqnarray}
where, $\tau$ is the optical path delay, $\omega$ is the angular frequency, and
\begin{equation}\label{eq1}
I(\tau ) = \frac{1}{{\pi (\Delta \omega )^2 }}\rm{exp}[ - \frac{1}{2}(\Delta \omega \tau )^2 ]
\end{equation}
is the indistinguishability \cite{Ra2013PNAS}, with $\Delta \omega$ corresponding to the spectral width of the photon source.
See the \textbf{Supplementary Information} for more details about the deduction of these equations, and the $P_{mn}$ for six-photon detection schemes.

With the experimentally measured bandwidth in the $1/1$ and $2/0$ schemes in Table \ref{table1}, we can estimate the $\Delta \omega$ value in Eq.\,(\ref{eq1}).
With the $\Delta \omega$ value and the equations of $P_{mn}$, we plot the theoretically expected interference patterns in Fig.\,\ref{results}(a2-j2).
It was interesting to notice that the two-photon schemes ($1/1$ and $2/0$ detection) have the same profiles, i.e. the profiles are detection-independent. However, for the four-photon  and six-photon schemes, the theoretical patterns are completely detection-dependent, i.e. different $m/n$ detection schemes have different profiles.
The visibilities for all the simulated patterns in Fig.\,\ref{results}(a2-j2) are normalized.
The main figures and insets in Fig.\,\ref{results}(a2-f2) correspond well with the experimental results in Fig.\,\ref{results}(a1-f1).
However, for the six-photon cases in Fig.\,\ref{results}(g2-j2)  and Fig.\,\ref{results}(g1-j1), there are several discrepancies for both the main figures and the insets. This was mainly caused by the strong multi-pair emission attributable a high pump power of 400~mW. The effect of multi-pair emission will be discussed in detail in the following section.

\textbf{\emph{Discussion}}
\noindent
%
%
\begin{figure*}[tbp]
\centering
\includegraphics[width= 0.95\textwidth]{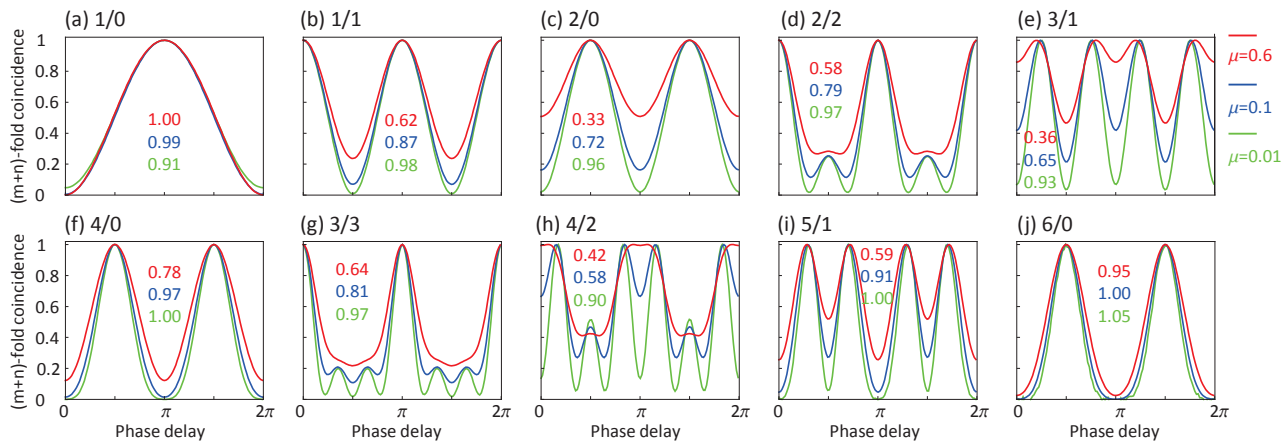}
\caption{ (color online)  Analysis of multi-pair emission. Theoretical multi-photon NOON-SI patterns with mean photon numbers of 0.01 (green), 0.1 (blue) and 0.6 (red). The visibilities are also labeled in the figure.
}
\label{multipair}
\end{figure*}
%
%
The multi-pair emission is an important reason for the  discrepancy between the theoretical simulations and the experimental results in Fig.\,\ref{results}. To obtain a reasonable counts for six photon detections in a short time, we applied a high pump power of 400 mW, which inevitably induced higher-order emissions. The multi-pair component in our PPKTP source as a function of pump power were investigated in Ref. \cite{Jin2015OC}. The mean photon numbers are about 0.1 and 0.6 for the pump power of 100 mW and 400 mW respectively. To analyze the effect of multi-pair emission in the interference patterns, we constructed a model of our experiment  using the characteristic functions method, which was previously used in entanglement swapping analysis  \cite{Jin2015SR, Takeoka2015NJP}. In this model, the SPDC source is considered as a squeezed vacuum and all the multi-pair components are included. Further, the collection efficiency $\eta$, dark counts $dc$ and mean photon numbers $\mu$ are also included in our model. The coincidence probability $p_{mn}$ can be written as
\begin{equation}\label{eq2}
p_{mn} = p_{mn}(\mu, \eta, dc, \phi)
\end{equation}
where $\mu$ is the mean photon number per pulse (also the squeezing parameter), $\eta$ is the total efficiency (the product of collection efficiency and detector efficiency), $dc$ is the dark count probability for one counting event, and $\phi$ is the phase delay.
With $\eta=0.2$, $dc=0.0001$, we plot the $p_{mn}$ as a function of the phase delay $\phi$ for  $\mu$ value of $0.01$, $0.1$, $0.6$, as shown in  Fig.\,\ref{multipair}(a-j).

In Fig.\,\ref{multipair}, it was interesting to notice that the $4/0$ and   $6/0$  schemes always keep relatively high visibilities, even for high mean photon numbers. This phenomenon is also verified by the experimental results in Fig.\,\ref{results}.
In contrast, the visibilities of the  $3/1$ and $4/2$ schemes decrease dramatically for higher mean photon numbers.
Further, the higher resolution fringes in the $3/1$ and $3/3$ schemes (which are 4 times finer than those of the $1/0$ scheme and 6 times finer, respectively) are very important for  high-precision phase sensing \cite{Nagata2007, Xiang2013}. However, these fine curves degrade rapidly at high $\mu$ values. Therefore, it is very important to maintain a low $\mu$ values for phase sensing applications of NOON state.

In Fig.\,\ref{results}, under the same pump power, the $2/0$ scheme has a higher visibility than the $1/1$ scheme. This result is different from our theoretical expectation in In Fig.\,\ref{multipair}. This discrepancy may be caused by the spatial mode matching condition in the experiment. To solve this problem in the future, more parameters should be included in the theoretical model.

It is also possible to extend our scheme to 8-photon or even 10-photon NOON-SI, if more efficient sources and detectors are available.
Especially, it is necessary to decrease the multi-photon emission, and one possible to way is to increase the repetition rate of the pump laser \cite{Jin2015SR}, which can decrease the multi-pair components without decreasing the count rates.

In the future, this experiment can be used for applications which are based on the envelope of the  NOON-SI pattern, such as quantum spectroscopy \cite{Dinani2016, Whittaker2015} and quantum metrology \cite{Dowling2008, Taylor2016}.

\textbf{\emph{Conclusion}}
\noindent
We experimentally and theoretically demonstrated six-, four- two- and one-photon NOON-SI with a spontaneous parametric down conversion source and six superconducting nanowire single-photon detectors. It was found that the shape, the coherence time and the visibility of the interference patterns are strongly dependent on the detection schemes. This is the first experimental observation of detection dependency of NOON-SI up to 6 photons at telecom wavelengths. This experiment can be used for applications which are based on the envelope of the NOON-SI pattern.

\textbf{\emph{Acknowledgements}}
The authors are grateful to T. Gerrits and A. Fedrizzi for insightful discussions, and to K. Wakui for assistance in experiment. This work was funded by ImPACT Program of Council for Science, Technology and Innovation (Cabinet Office, Government of Japan). R. J. Collins acknowledges funding from The Daiwa-Anglo Japanese Foundation through grant number 10803/11543.


\begin{thebibliography}{27}
\expandafter\ifx\csname natexlab\endcsname\relax\def\natexlab#1{#1}\fi
\expandafter\ifx\csname bibnamefont\endcsname\relax
  \def\bibnamefont#1{#1}\fi
\expandafter\ifx\csname bibfnamefont\endcsname\relax
  \def\bibfnamefont#1{#1}\fi
\expandafter\ifx\csname citenamefont\endcsname\relax
  \def\citenamefont#1{#1}\fi
\expandafter\ifx\csname url\endcsname\relax
  \def\url#1{\texttt{#1}}\fi
\expandafter\ifx\csname urlprefix\endcsname\relax\def\urlprefix{URL }\fi
\providecommand{\bibinfo}[2]{#2}
\providecommand{\eprint}[2][]{\url{#2}}

\bibitem[{\citenamefont{Boto et~al.}(2000)\citenamefont{Boto, Kok, Abrams,
  Braunstein, Williams, and Dowling}}]{Boto2000}
\bibinfo{author}{\bibfnamefont{A.~N.} \bibnamefont{Boto}},
  \bibinfo{author}{\bibfnamefont{P.}~\bibnamefont{Kok}},
  \bibinfo{author}{\bibfnamefont{D.~S.} \bibnamefont{Abrams}},
  \bibinfo{author}{\bibfnamefont{S.~L.} \bibnamefont{Braunstein}},
  \bibinfo{author}{\bibfnamefont{C.~P.} \bibnamefont{Williams}},
  \bibnamefont{and} \bibinfo{author}{\bibfnamefont{J.~P.}
  \bibnamefont{Dowling}}, \bibinfo{journal}{Phys. Rev. Lett.}
  \textbf{\bibinfo{volume}{85}}, \bibinfo{pages}{2733} (\bibinfo{year}{2000}).

\bibitem[{\citenamefont{Giovannetti et~al.}(2011)\citenamefont{Giovannetti,
  Lloyd, and Maccone}}]{Giovannetti2011}
\bibinfo{author}{\bibfnamefont{V.}~\bibnamefont{Giovannetti}},
  \bibinfo{author}{\bibfnamefont{S.}~\bibnamefont{Lloyd}}, \bibnamefont{and}
  \bibinfo{author}{\bibfnamefont{L.}~\bibnamefont{Maccone}},
  \bibinfo{journal}{Nat. Photon.} \textbf{\bibinfo{volume}{5}},
  \bibinfo{pages}{222} (\bibinfo{year}{2011}).

\bibitem[{\citenamefont{Ono et~al.}(2013)\citenamefont{Ono, Okamoto, and
  Takeuchi}}]{Ono2013}
\bibinfo{author}{\bibfnamefont{T.}~\bibnamefont{Ono}},
  \bibinfo{author}{\bibfnamefont{R.}~\bibnamefont{Okamoto}}, \bibnamefont{and}
  \bibinfo{author}{\bibfnamefont{S.}~\bibnamefont{Takeuchi}},
  \bibinfo{journal}{Nat. Commun.} \textbf{\bibinfo{volume}{4}},
  \bibinfo{pages}{2426} (\bibinfo{year}{2013}).

\bibitem[{\citenamefont{Israel et~al.}(2014)\citenamefont{Israel, Rosen, and
  Silberberg}}]{Israel2014}
\bibinfo{author}{\bibfnamefont{Y.}~\bibnamefont{Israel}},
  \bibinfo{author}{\bibfnamefont{S.}~\bibnamefont{Rosen}}, \bibnamefont{and}
  \bibinfo{author}{\bibfnamefont{Y.}~\bibnamefont{Silberberg}},
  \bibinfo{journal}{Phys. Rev. Lett.} \textbf{\bibinfo{volume}{112}},
  \bibinfo{pages}{103604} (\bibinfo{year}{2014}).

\bibitem[{\citenamefont{Bergmann and Loock}(2015)}]{Bergmann2015}
\bibinfo{author}{\bibfnamefont{M.}~\bibnamefont{Bergmann}} \bibnamefont{and}
  \bibinfo{author}{\bibfnamefont{P.~v.} \bibnamefont{Loock}},
  \bibinfo{journal}{arXiv:1512.07605}  (\bibinfo{year}{2015}).

\bibitem[{\citenamefont{Edamatsu et~al.}(2002)\citenamefont{Edamatsu, Shimizu,
  and Itoh}}]{Edamatsu2002}
\bibinfo{author}{\bibfnamefont{K.}~\bibnamefont{Edamatsu}},
  \bibinfo{author}{\bibfnamefont{R.}~\bibnamefont{Shimizu}}, \bibnamefont{and}
  \bibinfo{author}{\bibfnamefont{T.}~\bibnamefont{Itoh}},
  \bibinfo{journal}{Phys. Rev. Lett.} \textbf{\bibinfo{volume}{89}},
  \bibinfo{pages}{213601} (\bibinfo{year}{2002}).

\bibitem[{\citenamefont{Mitchell et~al.}(2004)\citenamefont{Mitchell, Lundeen,
  and Steinberg}}]{Mitchell2004}
\bibinfo{author}{\bibfnamefont{M.~W.} \bibnamefont{Mitchell}},
  \bibinfo{author}{\bibfnamefont{J.~S.} \bibnamefont{Lundeen}},
  \bibnamefont{and} \bibinfo{author}{\bibfnamefont{A.~M.}
  \bibnamefont{Steinberg}}, \bibinfo{journal}{Nature}
  \textbf{\bibinfo{volume}{429}}, \bibinfo{pages}{161} (\bibinfo{year}{2004}).

\bibitem[{\citenamefont{Walther et~al.}(2004)\citenamefont{Walther, Pan,
  Aspelmeyer, Ursin, Gasparoni, and Zeilinger}}]{Walther2004}
\bibinfo{author}{\bibfnamefont{P.}~\bibnamefont{Walther}},
  \bibinfo{author}{\bibfnamefont{J.-W.} \bibnamefont{Pan}},
  \bibinfo{author}{\bibfnamefont{M.}~\bibnamefont{Aspelmeyer}},
  \bibinfo{author}{\bibfnamefont{R.}~\bibnamefont{Ursin}},
  \bibinfo{author}{\bibfnamefont{S.}~\bibnamefont{Gasparoni}},
  \bibnamefont{and}
  \bibinfo{author}{\bibfnamefont{A.}~\bibnamefont{Zeilinger}},
  \bibinfo{journal}{Nature} \textbf{\bibinfo{volume}{429}},
  \bibinfo{pages}{158} (\bibinfo{year}{2004}).

\bibitem[{\citenamefont{Nagata et~al.}(2007)\citenamefont{Nagata, Okamoto,
  O'Brien, Sasaki, and Takeuchi}}]{Nagata2007}
\bibinfo{author}{\bibfnamefont{T.}~\bibnamefont{Nagata}},
  \bibinfo{author}{\bibfnamefont{R.}~\bibnamefont{Okamoto}},
  \bibinfo{author}{\bibfnamefont{J.~L.} \bibnamefont{O'Brien}},
  \bibinfo{author}{\bibfnamefont{K.}~\bibnamefont{Sasaki}}, \bibnamefont{and}
  \bibinfo{author}{\bibfnamefont{S.}~\bibnamefont{Takeuchi}},
  \bibinfo{journal}{Science} \textbf{\bibinfo{volume}{316}},
  \bibinfo{pages}{726} (\bibinfo{year}{2007}).

\bibitem[{\citenamefont{Afek et~al.}(2010)\citenamefont{Afek, Ambar, and
  Silberberg}}]{Afek2010}
\bibinfo{author}{\bibfnamefont{I.}~\bibnamefont{Afek}},
  \bibinfo{author}{\bibfnamefont{O.}~\bibnamefont{Ambar}}, \bibnamefont{and}
  \bibinfo{author}{\bibfnamefont{Y.}~\bibnamefont{Silberberg}},
  \bibinfo{journal}{Science} \textbf{\bibinfo{volume}{328}},
  \bibinfo{pages}{879} (\bibinfo{year}{2010}).

\bibitem[{\citenamefont{Xiang et~al.}(2013)\citenamefont{Xiang, Hofmann, and
  Pryde}}]{Xiang2013}
\bibinfo{author}{\bibfnamefont{G.~Y.} \bibnamefont{Xiang}},
  \bibinfo{author}{\bibfnamefont{H.~F.} \bibnamefont{Hofmann}},
  \bibnamefont{and} \bibinfo{author}{\bibfnamefont{G.~J.} \bibnamefont{Pryde}},
  \bibinfo{journal}{Sci. Rep.} \textbf{\bibinfo{volume}{3}},
  \bibinfo{pages}{2684} (\bibinfo{year}{2013}).

\bibitem[{\citenamefont{Yabuno et~al.}(2012)\citenamefont{Yabuno, Shimizu,
  Mitsumori, Kosaka, and Edamatsu}}]{Yabuno2012}
\bibinfo{author}{\bibfnamefont{M.}~\bibnamefont{Yabuno}},
  \bibinfo{author}{\bibfnamefont{R.}~\bibnamefont{Shimizu}},
  \bibinfo{author}{\bibfnamefont{Y.}~\bibnamefont{Mitsumori}},
  \bibinfo{author}{\bibfnamefont{H.}~\bibnamefont{Kosaka}}, \bibnamefont{and}
  \bibinfo{author}{\bibfnamefont{K.}~\bibnamefont{Edamatsu}},
  \bibinfo{journal}{Phys. Rev. A} \textbf{\bibinfo{volume}{86}},
  \bibinfo{pages}{010302} (\bibinfo{year}{2012}).

\bibitem[{\citenamefont{Bisht and Shimizu}(2015)}]{Bisht2015}
\bibinfo{author}{\bibfnamefont{N.~S.} \bibnamefont{Bisht}} \bibnamefont{and}
  \bibinfo{author}{\bibfnamefont{R.}~\bibnamefont{Shimizu}},
  \bibinfo{journal}{J. Opt. Soc. Am. B} \textbf{\bibinfo{volume}{32}},
  \bibinfo{pages}{550} (\bibinfo{year}{2015}).

\bibitem[{\citenamefont{Dinani et~al.}(2016)\citenamefont{Dinani, Gupta,
  Dowling, and Berry}}]{Dinani2016}
\bibinfo{author}{\bibfnamefont{H.~T.} \bibnamefont{Dinani}},
  \bibinfo{author}{\bibfnamefont{M.~K.} \bibnamefont{Gupta}},
  \bibinfo{author}{\bibfnamefont{J.~P.} \bibnamefont{Dowling}},
  \bibnamefont{and} \bibinfo{author}{\bibfnamefont{D.~W.} \bibnamefont{Berry}},
  \bibinfo{journal}{arXiv:1603.04509}  (\bibinfo{year}{2016}).

\bibitem[{\citenamefont{Whittaker et~al.}(2015)\citenamefont{Whittaker, Erven,
  Neville, Berry, O'Brien, Cable, and Matthews}}]{Whittaker2015}
\bibinfo{author}{\bibfnamefont{R.}~\bibnamefont{Whittaker}},
  \bibinfo{author}{\bibfnamefont{C.}~\bibnamefont{Erven}},
  \bibinfo{author}{\bibfnamefont{A.}~\bibnamefont{Neville}},
  \bibinfo{author}{\bibfnamefont{M.}~\bibnamefont{Berry}},
  \bibinfo{author}{\bibfnamefont{J.~L.} \bibnamefont{O'Brien}},
  \bibinfo{author}{\bibfnamefont{H.}~\bibnamefont{Cable}}, \bibnamefont{and}
  \bibinfo{author}{\bibfnamefont{J.~C.~F.} \bibnamefont{Matthews}},
  \bibinfo{journal}{arXiv:1508.00849}  (\bibinfo{year}{2015}).

\bibitem[{\citenamefont{Dowling}(2008)}]{Dowling2008}
\bibinfo{author}{\bibfnamefont{J.~P.} \bibnamefont{Dowling}},
  \bibinfo{journal}{Contemp. Phys.} \textbf{\bibinfo{volume}{49}},
  \bibinfo{pages}{125} (\bibinfo{year}{2008}).

\bibitem[{\citenamefont{Taylor and Bowen}(2016)}]{Taylor2016}
\bibinfo{author}{\bibfnamefont{M.~A.} \bibnamefont{Taylor}} \bibnamefont{and}
  \bibinfo{author}{\bibfnamefont{W.~P.} \bibnamefont{Bowen}},
  \bibinfo{journal}{Phys. Rep.} \textbf{\bibinfo{volume}{615}},
  \bibinfo{pages}{1} (\bibinfo{year}{2016}).

\bibitem[{\citenamefont{K{\"o}nig and Wong}(2004)}]{Konig2004}
\bibinfo{author}{\bibfnamefont{F.}~\bibnamefont{K{\"o}nig}} \bibnamefont{and}
  \bibinfo{author}{\bibfnamefont{F.~N.~C.} \bibnamefont{Wong}},
  \bibinfo{journal}{Appl. Phys. Lett.} \textbf{\bibinfo{volume}{84}},
  \bibinfo{pages}{1644} (\bibinfo{year}{2004}).

\bibitem[{\citenamefont{Jin et~al.}(2013)\citenamefont{Jin, Shimizu, Wakui,
  Benichi, and Sasaki}}]{Jin2013OE}
\bibinfo{author}{\bibfnamefont{R.-B.} \bibnamefont{Jin}},
  \bibinfo{author}{\bibfnamefont{R.}~\bibnamefont{Shimizu}},
  \bibinfo{author}{\bibfnamefont{K.}~\bibnamefont{Wakui}},
  \bibinfo{author}{\bibfnamefont{H.}~\bibnamefont{Benichi}}, \bibnamefont{and}
  \bibinfo{author}{\bibfnamefont{M.}~\bibnamefont{Sasaki}},
  \bibinfo{journal}{Opt. Express} \textbf{\bibinfo{volume}{21}},
  \bibinfo{pages}{10659} (\bibinfo{year}{2013}).

\bibitem[{\citenamefont{Miki et~al.}(2007)\citenamefont{Miki, Fujiwara, Sasaki,
  and Wang}}]{Miki2007}
\bibinfo{author}{\bibfnamefont{S.}~\bibnamefont{Miki}},
  \bibinfo{author}{\bibfnamefont{M.}~\bibnamefont{Fujiwara}},
  \bibinfo{author}{\bibfnamefont{M.}~\bibnamefont{Sasaki}}, \bibnamefont{and}
  \bibinfo{author}{\bibfnamefont{Z.}~\bibnamefont{Wang}},
  \bibinfo{journal}{IEEE Trans. Appl. Superconduct.}
  \textbf{\bibinfo{volume}{17}}, \bibinfo{pages}{285} (\bibinfo{year}{2007}).

\bibitem[{\citenamefont{Miki et~al.}(2013)\citenamefont{Miki, Yamashita, Terai,
  and Wang}}]{Miki2013}
\bibinfo{author}{\bibfnamefont{S.}~\bibnamefont{Miki}},
  \bibinfo{author}{\bibfnamefont{T.}~\bibnamefont{Yamashita}},
  \bibinfo{author}{\bibfnamefont{H.}~\bibnamefont{Terai}}, \bibnamefont{and}
  \bibinfo{author}{\bibfnamefont{Z.}~\bibnamefont{Wang}},
  \bibinfo{journal}{Opt. Express} \textbf{\bibinfo{volume}{21}},
  \bibinfo{pages}{10208} (\bibinfo{year}{2013}).

\bibitem[{\citenamefont{Jin et~al.}(2015{\natexlab{a}})\citenamefont{Jin,
  Fujiwara, Yamashita, Miki, Terai, Wang, Wakui, Shimizu, and
  Sasaki}}]{Jin2015OC}
\bibinfo{author}{\bibfnamefont{R.-B.} \bibnamefont{Jin}},
  \bibinfo{author}{\bibfnamefont{M.}~\bibnamefont{Fujiwara}},
  \bibinfo{author}{\bibfnamefont{T.}~\bibnamefont{Yamashita}},
  \bibinfo{author}{\bibfnamefont{S.}~\bibnamefont{Miki}},
  \bibinfo{author}{\bibfnamefont{H.}~\bibnamefont{Terai}},
  \bibinfo{author}{\bibfnamefont{Z.}~\bibnamefont{Wang}},
  \bibinfo{author}{\bibfnamefont{K.}~\bibnamefont{Wakui}},
  \bibinfo{author}{\bibfnamefont{R.}~\bibnamefont{Shimizu}}, \bibnamefont{and}
  \bibinfo{author}{\bibfnamefont{M.}~\bibnamefont{Sasaki}},
  \bibinfo{journal}{Opt. Commun.} \textbf{\bibinfo{volume}{336}},
  \bibinfo{pages}{47} (\bibinfo{year}{2015}{\natexlab{a}}).

\bibitem[{\citenamefont{Tichy et~al.}(2011)\citenamefont{Tichy, Lim, Ra,
  Mintert, Kim, and Buchleitner}}]{Tichy2011}
\bibinfo{author}{\bibfnamefont{M.~C.} \bibnamefont{Tichy}},
  \bibinfo{author}{\bibfnamefont{H.-T.} \bibnamefont{Lim}},
  \bibinfo{author}{\bibfnamefont{Y.-S.} \bibnamefont{Ra}},
  \bibinfo{author}{\bibfnamefont{F.}~\bibnamefont{Mintert}},
  \bibinfo{author}{\bibfnamefont{Y.-H.} \bibnamefont{Kim}}, \bibnamefont{and}
  \bibinfo{author}{\bibfnamefont{A.}~\bibnamefont{Buchleitner}},
  \bibinfo{journal}{Phys. Rev. A} \textbf{\bibinfo{volume}{83}},
  \bibinfo{pages}{062111} (\bibinfo{year}{2011}).

\bibitem[{\citenamefont{Ra et~al.}(2013{\natexlab{a}})\citenamefont{Ra, Tichy,
  Lim, Kwon, Mintert, Buchleitner, and Kim}}]{Ra2013NC}
\bibinfo{author}{\bibfnamefont{Y.-S.} \bibnamefont{Ra}},
  \bibinfo{author}{\bibfnamefont{M.~C.} \bibnamefont{Tichy}},
  \bibinfo{author}{\bibfnamefont{H.-T.} \bibnamefont{Lim}},
  \bibinfo{author}{\bibfnamefont{O.}~\bibnamefont{Kwon}},
  \bibinfo{author}{\bibfnamefont{F.}~\bibnamefont{Mintert}},
  \bibinfo{author}{\bibfnamefont{A.}~\bibnamefont{Buchleitner}},
  \bibnamefont{and} \bibinfo{author}{\bibfnamefont{Y.-H.} \bibnamefont{Kim}},
  \bibinfo{journal}{Nat. Commun.} \textbf{\bibinfo{volume}{4}},
  \bibinfo{pages}{2451} (\bibinfo{year}{2013}{\natexlab{a}}).

\bibitem[{\citenamefont{Ra et~al.}(2013{\natexlab{b}})\citenamefont{Ra, Tichy,
  Lim, Kwon, Mintert, Buchleitner, and Kim}}]{Ra2013PNAS}
\bibinfo{author}{\bibfnamefont{Y.-S.} \bibnamefont{Ra}},
  \bibinfo{author}{\bibfnamefont{M.~C.} \bibnamefont{Tichy}},
  \bibinfo{author}{\bibfnamefont{H.-T.} \bibnamefont{Lim}},
  \bibinfo{author}{\bibfnamefont{O.}~\bibnamefont{Kwon}},
  \bibinfo{author}{\bibfnamefont{F.}~\bibnamefont{Mintert}},
  \bibinfo{author}{\bibfnamefont{A.}~\bibnamefont{Buchleitner}},
  \bibnamefont{and} \bibinfo{author}{\bibfnamefont{Y.-H.} \bibnamefont{Kim}},
  \bibinfo{journal}{PNAS} \textbf{\bibinfo{volume}{110}}, \bibinfo{pages}{1227}
  (\bibinfo{year}{2013}{\natexlab{b}}).

\bibitem[{\citenamefont{Jin et~al.}(2015{\natexlab{b}})\citenamefont{Jin,
  Takeoka, Takagi, Shimizu, and Sasaki}}]{Jin2015SR}
\bibinfo{author}{\bibfnamefont{R.-B.} \bibnamefont{Jin}},
  \bibinfo{author}{\bibfnamefont{M.}~\bibnamefont{Takeoka}},
  \bibinfo{author}{\bibfnamefont{U.}~\bibnamefont{Takagi}},
  \bibinfo{author}{\bibfnamefont{R.}~\bibnamefont{Shimizu}}, \bibnamefont{and}
  \bibinfo{author}{\bibfnamefont{M.}~\bibnamefont{Sasaki}},
  \bibinfo{journal}{Sci. Rep.} \textbf{\bibinfo{volume}{5}},
  \bibinfo{pages}{9333} (\bibinfo{year}{2015}{\natexlab{b}}).

\bibitem[{\citenamefont{Takeoka et~al.}(2015)\citenamefont{Takeoka, Jin, and
  Sasaki}}]{Takeoka2015NJP}
\bibinfo{author}{\bibfnamefont{M.}~\bibnamefont{Takeoka}},
  \bibinfo{author}{\bibfnamefont{R.-B.} \bibnamefont{Jin}}, \bibnamefont{and}
  \bibinfo{author}{\bibfnamefont{M.}~\bibnamefont{Sasaki}},
  \bibinfo{journal}{New J. Phys.} \textbf{\bibinfo{volume}{17}},
  \bibinfo{pages}{043030} (\bibinfo{year}{2015}).

\end{thebibliography}

\end{document}